\documentclass[twocolumn,pra,showpacs]{revtex4}
\usepackage{amssymb}
\usepackage{amsfonts}
\usepackage{amsmath}
\usepackage{graphicx}

\begin{document}

\title{Markovian embedding of non-Markovian quantum collisional models}
\author{Adri\'{a}n A. Budini}
\affiliation{Consejo Nacional de Investigaciones Cient\'{\i}ficas y T\'{e}cnicas
(CONICET), Centro At\'{o}mico Bariloche, Avenida E. Bustillo Km 9.5, (8400)
Bariloche, Argentina,}
\affiliation{Universidad Tecnol\'{o}gica Nacional (UTN-FRBA), Fanny Newbery 111, (8400)
Bariloche, Argentina}
\date{\today }

\begin{abstract}
A wide class of non-Markovian completely positive master equations can be
formulated on the basis of quantum collisional models. In this
phenomenological approach the dynamics of an open quantum system is modeled
through an ensemble of stochastic realizations that consist in the
application at random times of a (collisional) completely positive
transformation over the system state. In this paper, we demonstrate that
these kinds of models can be embedded in bipartite Markovian Lindblad
dynamics consisting of the system of interest and an auxiliary one. In
contrast with phenomenological formulations, here the stochastic ensemble
dynamics an the inter-event time interval statistics are obtained from a
quantum measurement theory after assuming that the auxiliary system is
continuously monitored in time. Models where the system inter-collisional
dynamics is non-Markovian [B. Vacchini, Phys. Rev. A \textbf{87}, 030101(R)
(2013)] are also obtained from the present approach. The formalism is
exemplified through bipartite dynamics that leads to non-Markovian system
effects such as an environment-to-system back flow of information.
\end{abstract}

\pacs{03.65.Yz, 42.50.Lc, 03.65.Ta, 02.50.Ga}
\maketitle

%03.65.Yz Decoherence open systems
%42.50.Lc Quantum fluctuations, quantum noise, and quantum jumps
%03.65.Ta Foundations of quantum mechanics; measurement theory
%02.50.Ga Markov processes

%33.80.-b Photon interactions with molecules
%42.50.Ar Photon statistics and coherence theory

\section{Introduction}

The description of open quantum systems through local in time Markovian
evolutions is well understood from both mathematical \cite{alicki} and
physical \cite{breuerbook} point of views. As is well know, under a
completely positive condition, Lindblad equations provide the more general
evolution structure of the system density matrix \cite{alicki,breuerbook}.
On the other hand, in the last years an ever increasing interest have been
paid to establishing a non-Markovian generalization of the open quantum
system theory formulated in terms of non-local in time evolutions \cite%
{haake}. %In general, one can link the
%memory contributions with the capability of monitoring and controlling the
%time-irreversible influence of the environment degrees of freedom. Hence,
%besides its pure theoretical interest, a non-Markovian generalization of the
%theory of open quantum systems may play a fundamental role in describing
%coherent quantum features of engineered system-bath interactions.
There exist diverse formalisms for describing memory effects. One leading
program consist in generalizing Lindblad equations by replacing the rates of
each dissipative channel by a time-convoluted kernel function. A wide class
of both phenomenological \cite{stenholm,classBu,GrigoBu,VacCol,giraldi} and
theoretical approaches \cite%
{wilkie,salo,Lidar,cresser,ManisPetru,rate,vacchini,SemiMarkov,Kosa,maniscalco,BuJumpSMS,petruccione,barchielli,OneChannel}
were formulated for building and characterizing master equations of that
kind, which in turn lead to a completely positive solution map.

In the category of phenomenological approaches, quantum collisional models
(QCMs) provided a fundamental tool for establishing a non-Markovian
generalization of Lindblad equations \cite{stenholm,classBu}. In this
formalism, the evolution of an open quantum system follows from an average
performed over an ensemble of stochastic realizations of the system state.
Each realization consists in the application, at random times, of a
completely positive transformation. The events can be read as a
\textquotedblleft collision\textquotedblright\ or interaction with the
environment. Depending on the statistics of the collision times and the
system inter-event dynamics different non-Markovian master equations were
established \cite{classBu,GrigoBu,VacCol,giraldi}. Over that basis, the
emergence of non-Markovian effects such as a system-to-environment back flow
of information \cite{NoMeasure,Entropy} were also analyzed in the recent
literature \cite{pheno,pheno1}.

The collisional superoperator, the inter-event system dynamics, and the
collision time statistics are the main ingredients of the approach. They
must be defined, in an arbitrary way, from the beginning. Therefore, besides
its usefulness, the QCM model does not have associated a microscopic
description, neither it is completely understood which kind of underlying
mechanism may induce the structure of the stochastic dynamics. The main goal
of this paper is to provide a rigorous physical frame to\ answering these
issues.

The basic idea consists in embedding the non-Markovian system evolution in a
Markovian bipartite dynamics. It is defined by the system of interest and an
auxiliary (ancilla) system. We demonstrate that there exist bipartite
Markovian interactions that induce the same system non-Markovian dynamics.
In this way, \textquotedblleft microscopic interactions\textquotedblright\
that lead to the master equations associated to the QCM are found. On the
other hand, by assuming that the auxiliary system in continuously monitored
in time, over the basis of a (Markovian) quantum jump approach \cite%
{plenio,carmichaelbook,carmichael}, we find that the realizations of the QCM
can be put in one to one correspondence with the realizations of the
measurement apparatus. In this way, the stochastic dynamics of the QCM is
established from a quantum measurement theory. In addition, this modeling
allows to characterizing the inter-event statistics from the Markovian
Lindblad description.

In Ref. \cite{VacCol} Vacchini introduced a generalized QCM where, in
contrast to previous approaches \cite{classBu,GrigoBu,giraldi}, the system
inter-event dynamics is defined by a non-Markovian propagator. On the basis
of an underlying tripartite Markovian dynamics we show that this
generalization can also be described with the present frame. Even when the
stochastic realizations consist of successive collisional events with a
non-Markovian inter-event dynamics \cite{VacCol}, they cannot be read as the
result of a continuous measurement action performed over the system of
interest. In fact, in contrast with the results of Ref. \cite{OneChannel},
here we demonstrate that QCMs can consistently be recovered when measuring
the auxiliary ancilla system. The non-Markovian quantum jump approach
developed in \cite{OneChannel} relies on more general bipartite
interactions. Additionally, the monitoring action is performed over the
system of interest. 
%On the other hand, by working explicitly the non-Markovian dynamics of a
%two-level system, consistently with the analyisis of Ref. \cite{pheno}, we
%show that collisional models can also lead to non-Markovian effects such as
%an environment-to-system back flow of information. The emergence of this
%property, which has been extensively analyzed in the recent literature \cite%
%{NoMeasure}, as well as the present formulation demonstrate that collisional
%models provide a valid tool for describing non-Markovian completely positive
%dynamics. In fact, we 

It is interesting to note that collisional models were also proposed as a
phenomenological tool for deriving Markovian irreversible dynamics \cite%
{alicki,rau}. Furthermore, from a quantum information perspective \cite%
{nielsen}, similar approaches were introduced by considering collisions with
a string of auxiliary qubits systems \cite{gisin,buzeMarkov,palma}. When the
system-string interaction is defined by partial swap and controlled-not
qubits operations, specific Markovian master equations describe the system
dynamics \cite{buzeMarkov}. Generalization of these ideas to non-Markovian
dynamics were considered recently in Refs. \cite{tanos,Giova,buzeta}. The
stretched relation of these results with the present formalism is also
investigated.

The paper is outlined as follows. In Sect. II we present the Markovian
embedding, where the system density matrix is obtained by using projector
techniques \cite{haake}. In Sect. III, from a standard quantum measurement
theory, we obtain the stochastic ensemble dynamic after assuming that the
auxiliary system is subjected to a measurement process. These results relies
in the standard quantum jump approach \cite{plenio,carmichaelbook,carmichael}
applied to bipartite dynamics. In Sec. IV, we analyze some examples that
exhibits the main features of the present approach. A back flow of
information from the system to the environment is explicitly shown. In Sec.
V, some generalizations of the standard collisional approach are provided.
The dynamics presented in Ref. \cite{VacCol} is recovered from a tripartite
Markovian dynamics. The formalisms of Refs. \cite{tanos,Giova,buzeta} are
analized in this context. In Sec. VI we present the conclusions.

\section{Markovian embedding}

In this section, it is demonstrated that non-Markovian QCMs can be obtained
by tracing out a bipartite Markovian dynamics. We deal the case of
stationary renewal statistics.

\subsection{Phenomenological renewal collisional models}

The superoperator $\mathcal{E}_{s}$\ that define each collisional events is
written\ as%
\begin{equation}
\mathcal{E}_{s}[\rho ]=\sum_{\alpha }V_{\alpha }\rho V_{\alpha }^{\dag },\ \
\ \ \ \ \ \ \sum_{\alpha }V_{\alpha }^{\dag }V_{\alpha }=\mathrm{I}_{s},
\label{Esuperoperator}
\end{equation}%
where the set of operators $\{V_{\alpha }\}$ act on the system Hilbert
space. $\mathrm{I}_{s}$ is the identity matrix. Between collision events the
system dynamics is defined by an arbitrary Lindblad generator $\mathcal{L}%
_{s}$ [unitary plus dissipative contributions]. Thus, given that the last
event happened at time $t^{\prime },$ the inter-event evolution follows from
the propagator $\exp [(t-t^{\prime })\mathcal{L}_{s}].$ By assuming that the
collision times define a renewal process, with waiting time distribution $%
w(t)$ \cite{classBu}, it is possible to demonstrate that the average system
density matrix $\rho _{t}^{s}$ is governed by the equation \cite{GrigoBu}%
\begin{equation}
\frac{d}{dt}\rho _{t}^{s}=\mathcal{L}_{s}[\rho
_{t}^{s}]+\int_{0}^{t}dt^{\prime }k(t-t^{\prime })\mathcal{C}_{s}\{\exp
[(t-t^{\prime })\mathcal{L}_{s}]\rho _{t^{\prime }}^{s}\}.
\label{MasterColision}
\end{equation}%
The superoperator $\mathcal{C}_{s}$ and the kernel function read%
\begin{equation}
\mathcal{C}_{s}=\mathcal{E}_{s}-\mathrm{I}_{s},\ \ \ \ \ \ \ \ \ \ \ k(u)=%
\frac{uw(u)}{1-w(u)},  \label{CollisionSystemWaiting}
\end{equation}%
where $u$ is a Laplace variable $[f(u)\equiv \int_{0}^{\infty
}dte^{-ut}f(t)].$ Notice that here, due to the assumed (stationary) renewal
property, the kernel does not depend separately on the time variables $t$
and $t^{\prime }.$ On the other hand, if $[\mathcal{C}_{s},\mathcal{L}%
_{s}]=0,$ in an interaction representation with respect to $\mathcal{L}_{s}$
Eq. (\ref{MasterColision}) (under the replacement $\mathcal{L}%
_{s}\rightarrow 0$) recovers the evolution introduced in Ref. \cite{classBu}.

\subsection{Bipartite Markovian dynamics}

We introduce a bipartite arrangement defined by the system of interest $S$
and an auxiliary (ancilla) system $A.$ Their joint density matrix is $\rho
_{t}^{sa}.$ Therefore, their marginal density matrices follow from a partial
trace,%
\begin{equation}
\rho _{t}^{s}=\mathrm{Tr}_{a}\left[ \rho _{t}^{sa}\right] ,\ \ \ \ \ \ \ \ \
\ \rho _{t}^{a}=\mathrm{Tr}_{s}\left[ \rho _{t}^{sa}\right] .  \label{RhoS_A}
\end{equation}%
The bipartite dynamics is defined by a Markovian Lindblad equation%
\begin{equation}
\frac{d}{dt}\rho _{t}^{sa}=\mathcal{L}\rho _{t}^{sa}=(\mathcal{L}_{s}+%
\mathcal{L}_{a}+\mathcal{C}_{sa})\rho _{t}^{sa},  \label{LindbladBipartito}
\end{equation}%
where\ the (arbitrary) Lindblad generators $\mathcal{L}_{s}$ and $\mathcal{L}%
_{a}$ define the system and ancilla dynamics respectively. The contribution $%
\mathcal{C}_{sa}$ introduces their mutual interaction.

Now we ask about the possibility of finding specific system-ancilla
interactions such that the marginal system density matrix $\rho _{t}^{s}$
[Eq. (\ref{RhoS_A})] fulfill the evolution (\ref{MasterColision}). With this
goal in mind, the superoperator $\mathcal{C}_{sa}$ is defined as%
\begin{equation}
\mathcal{C}_{sa}[\rho ]=\sum_{\alpha ,l}\gamma _{l}([V_{\alpha l},\rho
V_{\alpha l}^{\dag }]+[V_{\alpha l}\rho ,V_{\alpha l}^{\dag }]),
\label{LindbladInteraction}
\end{equation}%
where $\gamma _{l}$ are dissipative rates and the operator $V_{\alpha l}$\ is%
\begin{equation}
V_{\alpha l}=V_{\alpha }\otimes \left\vert a_{l}\right\rangle \left\langle
a_{\mathrm{0}}\right\vert .  \label{Operators}
\end{equation}%
The set of operators $\{V_{\alpha }\}$ are the same than in Eq. (\ref%
{Esuperoperator}). The states $\{\left\vert a_{\mathrm{0}}\right\rangle
,\left\vert a_{l}\right\rangle \},$ $l=1,2,\cdots ,\dim \{\mathcal{H}%
_{a}\}-1,$ form a complete orthogonal normalized basis in the Hilbert space $%
\mathcal{H}_{a}$ of the ancilla system. Hence, excepting the state $%
\left\vert a_{\mathrm{0}}\right\rangle ,$ the index $l$ runs over all
available states. Notice\ that operators (\ref{Operators}) introduce
irreversible ancilla transitions between the state $\left\vert a_{\mathrm{0}%
}\right\rangle $ and any of the remaining possible states $\left\vert
a_{l}\right\rangle ,$ that is, $\left\vert a_{\mathrm{0}}\right\rangle
\rightsquigarrow \left\vert a_{l}\right\rangle .$

\subsubsection{Ancilla dynamics}

With the previous choice of operators [Eq. (\ref{Operators})], it is simple
to write down a closed Markovian evolution for the ancilla state$\ \rho
_{t}^{a}.$ From Eqs. (\ref{LindbladBipartito}) and (\ref{LindbladInteraction}%
) we get%
\begin{equation}
\frac{d}{dt}\rho _{t}^{a}=\mathbb{L}_{a}\rho _{t}^{a}=(\mathcal{L}_{a}+%
\mathcal{C}_{a})\rho _{t}^{a}.  \label{AncillaEvolution}
\end{equation}%
The extra Lindblad term reads%
\begin{equation}
\mathcal{C}_{a}\rho _{t}^{a}=\sum_{l}\gamma _{l}([A_{l},\rho
_{t}^{a}A_{l}^{\dag }]+[A_{l}\rho _{t}^{a},A_{l}^{\dag }]),
\end{equation}%
where $A_{l}=\left\vert a_{l}\right\rangle \left\langle a_{\mathrm{0}%
}\right\vert .$ Straightforwardly, this superoperator can be rewritten as 
\begin{equation}
\mathcal{C}_{a}\rho _{t}^{a}=-\frac{1}{2}\gamma \{\left\vert a_{\mathrm{0}%
}\right\rangle \left\langle a_{\mathrm{0}}\right\vert ,\rho
_{t}^{a}\}_{+}+\gamma \left\langle a_{\mathrm{0}}\right\vert \rho
_{t}^{a}\left\vert a_{\mathrm{0}}\right\rangle \bar{\rho}_{a}.  \label{Ca}
\end{equation}%
Here, $\{\cdots \}_{+}$ denotes an anticommutation operation, and the
ancilla state $\bar{\rho}_{a}$ is%
\begin{equation}
\bar{\rho}_{a}=\sum_{l}\frac{\gamma _{l}}{\gamma }\left\vert
a_{l}\right\rangle \left\langle a_{l}\right\vert ,\ \ \ \ \ \ \ \ \ \gamma
=\sum_{l}\gamma _{l},  \label{AncillaReset}
\end{equation}%
which in fact satisfies $\mathrm{Tr}_{a}\left[ \bar{\rho}_{a}\right] =1.$

\subsubsection{Non-Markovian system dynamics}

In contrast to Eq. (\ref{AncillaEvolution}), the evolution of the system
state $\rho _{t}^{s}$ is non-Markovian. Its calculation is a little more
involved, which here is obtained by using a projector technique \cite%
{breuerbook,haake}. Let introduce the projectors $\mathcal{P}$ and $\mathcal{%
Q}$,%
\begin{equation}
\mathcal{P}\rho _{t}^{sa}=\mathrm{Tr}_{a}\left[ \rho _{t}^{sa}\right]
\otimes \bar{\rho}_{a},\ \ \ \ \ \ \ \ \mathcal{P}+\mathcal{Q}=\mathrm{I}%
_{sa},  \label{projectors}
\end{equation}%
where $\mathrm{I}_{sa}$ is the identity matrix in the bipartite
system-ancilla Hilbert space. $\bar{\rho}_{a}$\ is the ancilla state (\ref%
{AncillaReset}). The election of this projector definition will becomes
clear in the next section.

The bipartite evolution (\ref{LindbladBipartito}) can be projected in a
relevant and irrelevant contributions \cite{haake}%
\begin{eqnarray}
\frac{d}{dt}\mathcal{P}\rho _{t}^{sa} &=&\mathcal{PL}(\mathcal{P}+\mathcal{Q}%
)\rho _{t}^{sa},  \label{Relevante} \\
\frac{d}{dt}\mathcal{Q}\rho _{t}^{sa} &=&\mathcal{QL}(\mathcal{P}+\mathcal{Q}%
)\rho _{t}^{sa}.  \label{Irrelevante}
\end{eqnarray}%
On the other hand, consistently with the projectors definition (\ref%
{projectors}), a separable state defines the bipartite initial condition%
\begin{equation}
\rho _{0}^{sa}=\rho _{0}^{s}\otimes \bar{\rho}_{a},  \label{CISeparable}
\end{equation}%
where $\rho _{0}^{s}$ is an arbitrary system state. With this initial state,
it follows that $\mathcal{Q}\rho _{0}^{sa}=0.$ Therefore, Eq. (\ref%
{Irrelevante}) can be integrated \cite{haake} as $\mathcal{Q}\rho
_{t}^{sa}=\int_{0}^{t}dt^{\prime }\exp [\mathcal{QL}(t-t^{\prime })]\mathcal{%
QLP}\rho _{t^{\prime }}^{sa},$ which in turn, after replacing in Eq. (\ref%
{Relevante}) leads to the convoluted evolution%
\begin{equation}
\frac{d}{dt}\mathcal{P}\rho _{t}^{sa}=\mathcal{PLP}\rho _{t}^{sa}+\mathcal{PL%
}\int_{0}^{t}dt^{\prime }\exp [\mathcal{QL}(t-t^{\prime })]\mathcal{QLP}\rho
_{t^{\prime }}^{sa}.  \label{EvolutionProjectores}
\end{equation}%
The superoperator $\mathcal{L}$ is defined by Eq. (\ref{LindbladBipartito}).
From Eqs. (\ref{LindbladInteraction}) and (\ref{Operators}), it can be
rewritten as%
\begin{eqnarray}
\mathcal{L}[\bullet ] &=&(\mathcal{L}_{s}+\mathcal{L}_{a})[\bullet ]-\frac{1%
}{2}\gamma \{\left\vert a_{\mathrm{0}}\right\rangle \left\langle a_{\mathrm{0%
}}\right\vert ,\bullet \}_{+}  \notag \\
&&+\gamma \mathcal{E}_{s}[\left\langle a_{\mathrm{0}}\right\vert \bullet
\left\vert a_{\mathrm{0}}\right\rangle ]\otimes \bar{\rho}_{a},
\label{eleExplicito}
\end{eqnarray}%
where the collision superoperator $\mathcal{E}_{s}$ and the ancilla state $%
\bar{\rho}_{a}$ are defined by Eqs. (\ref{Esuperoperator}) and (\ref%
{AncillaReset})\ respectively. Eqs. (\ref{projectors}) and (\ref%
{eleExplicito}) lead to%
\begin{equation}
\mathcal{PL}[\bullet ]=\{\mathcal{L}_{s}(\mathrm{Tr}_{a}\left[ \bullet %
\right] )+\gamma \mathcal{C}_{s}[\left\langle a_{\mathrm{0}}\right\vert
\bullet \left\vert a_{\mathrm{0}}\right\rangle ]\}\otimes \bar{\rho}_{a},
\end{equation}%
where $\mathcal{C}_{s}$\ follows from Eq. (\ref{CollisionSystemWaiting}).
With these last two expressions it is possible to evaluate all contributions
in Eq. (\ref{EvolutionProjectores}). By using that $\left\langle a_{\mathrm{0%
}}\right\vert \bar{\rho}_{a}\left\vert a_{\mathrm{0}}\right\rangle =0,$ we
get $\mathcal{PLP}\rho _{t}^{sa}=\mathcal{L}_{s}[\rho _{t}^{s}]\otimes \bar{%
\rho}_{a},$ and $\mathcal{QLP}\rho _{t^{\prime }}^{sa}=\rho _{t^{\prime
}}^{s}\otimes \mathbb{L}_{a}[\bar{\rho}_{a}],$ where the ancilla
superoperator $\mathbb{L}_{a}$ follows from Eq. (\ref{AncillaEvolution}). We
have also used that $\mathcal{C}_{a}[\bar{\rho}_{a}]=0$ [see Eqs. (\ref{Ca})
and (\ref{AncillaReset})]. Similarly, it is possible to demonstrate that $%
\mathcal{QL}(\rho _{t^{\prime }}^{s}\otimes \mathbb{L}_{a}[\bar{\rho}_{a}])=(%
\mathcal{L}_{s}+\mathbb{L}_{a})(\rho _{t^{\prime }}^{s}\otimes \mathbb{L}%
_{a}[\bar{\rho}_{a}]),$ which by induction implies the expression%
\begin{equation}
\exp [\mathcal{QL}t]\mathcal{QLP}\rho _{t^{\prime }}^{sa}=\exp [(\mathcal{L}%
_{s}+\mathbb{L}_{a})t](\rho _{t^{\prime }}^{s}\otimes \mathbb{L}_{a}[\bar{%
\rho}_{a}]).
\end{equation}%
By introducing the previous results in Eq. (\ref{EvolutionProjectores}),
using that $\mathrm{Tr}_{a}[\mathbb{L}_{a}(\bullet )]=0,$ straightforwardly
we recover the convoluted evolution (\ref{MasterColision}) with the kernel
function 
\begin{subequations}
\label{LindbladKernel}
\begin{eqnarray}
k(t) &=&\gamma \left\langle a_{\mathrm{0}}\right\vert \exp (t\mathbb{L}_{a})%
\mathbb{L}_{a}[\bar{\rho}_{a}]\left\vert a_{\mathrm{0}}\right\rangle , \\
&=&\gamma \frac{d}{dt}\left\langle a_{\mathrm{0}}\right\vert \exp (t\mathbb{L%
}_{a})[\bar{\rho}_{a}]\left\vert a_{\mathrm{0}}\right\rangle .
\end{eqnarray}%
This is the main result of this section. It demonstrate that the
non-Markovian evolution (\ref{MasterColision}) also arises as the marginal
dynamics of a Markovian bipartite dynamics. In addition, here the kernel
function is not arbitrary. In fact, it is completely determined from the
ancilla dynamics [see Eqs. (\ref{AncillaEvolution}) and (\ref{LindbladKernel}%
)]. Notice that the solution map $\rho _{0}^{s}\rightarrow \rho _{t}^{s}$
associated to the evolution (\ref{MasterColision}) with the kernel (\ref%
{LindbladKernel}) is, by construction, completely positive.

\section{Quantum measurement theory}

In the previous section we have found an underlying bipartite Markovian
dynamics that leads to the\ non-Markovian system dynamics. Here, over the
same basis we find a clear physical interpretation to the ensemble of
realizations \cite{classBu,GrigoBu} associated to the master equation (\ref%
{MasterColision}).

\subsection{Quantum jumps in the bipartite dynamics}

The realizations of the collision model do not rely on a quantum measurement
theory. Nevertheless, this link can be established by studying the bipartite
dynamics when a measurement process is performed over the ancilla system.
Specifically, we assume that the apparatus is sensitive to all ancilla
transitions $\left\vert a_{\mathrm{0}}\right\rangle \rightsquigarrow
\left\vert a_{l}\right\rangle .$ As the bipartite dynamics is Markovian,
from a standard quantum jump approach \cite{plenio,carmichaelbook} it is
possible to associate each realization of the monitoring process with a
realization in the system-ancilla Hilbert space such that

\end{subequations}
\begin{equation}
\rho _{t}^{sa}=\overline{\rho _{sa}^{\mathrm{st}}(t)}.  \label{RhoAverage}
\end{equation}%
Here, $\rho _{sa}^{\mathrm{st}}(t)$ is a stochastic density matrix and the
overbar denotes an ensemble average. The time evolution of $\rho _{t}^{sa}$
is defined by Eq. (\ref{LindbladBipartito}). As usual, the stochastic
dynamics of $\rho _{sa}^{\mathrm{st}}(t)$\ consists of disruptive
transformations associated to each recording event, while in the
intermediate time intervals it is smooth and non-unitary \cite%
{plenio,carmichaelbook}.

Consistently with a quantum measurement theory \cite{breuerbook}, in each
detection event the bipartite state suffer the (measurement) transformation 
\begin{equation}
\rho \rightarrow \mathcal{M}\rho =\frac{\mathcal{J}\rho }{\mathrm{Tr}_{sa}[%
\mathcal{J}\rho ]},  \label{BipartiteMeasurement}
\end{equation}%
where the superoperator $\mathcal{J}$\ takes into account all possible
transitions $\left\vert a_{\mathrm{0}}\right\rangle \rightsquigarrow
\left\vert a_{l}\right\rangle $ that lead to a detection event. Assuming
that $\mathcal{L}_{a}$\ does not induce this kind of transitions, from Eq. (%
\ref{LindbladInteraction}) we write 
\begin{subequations}
\label{MBpartito}
\begin{eqnarray}
\mathcal{M}\rho &=&\frac{\sum_{\alpha l}\gamma _{l}V_{\alpha l}\rho
V_{\alpha l}^{\dag }}{\{\mathrm{Tr}_{sa}[\sum_{\alpha l}\gamma _{l}V_{\alpha
l}^{\dag }V_{\alpha l}\rho ]\}}, \\
&=&\frac{\mathcal{E}_{s}\left\langle a_{\mathrm{0}}\right\vert \rho
\left\vert a_{\mathrm{0}}\right\rangle }{\mathrm{Tr}_{s}[\left\langle a_{%
\mathrm{0}}\right\vert \rho \left\vert a_{\mathrm{0}}\right\rangle ]}\otimes 
\bar{\rho}_{a}.
\end{eqnarray}%
This last expression follows from the definition of the operators $V_{\alpha
l},$ Eq. (\ref{Operators}). On the other hand, the conditional evolution of $%
\rho _{sa}^{\mathrm{st}}(t)$ between detection events is given by the
normalized propagator \cite{plenio,carmichaelbook} 
\end{subequations}
\begin{equation}
\mathcal{T}_{c}(t)\rho =\frac{\mathcal{T}(t)\rho }{\mathrm{Tr}_{sa}[\mathcal{%
T}(t)\rho ]},  \label{Conditional}
\end{equation}%
where the unnormalized propagator $\mathcal{T}(t)$ is%
\begin{equation}
\mathcal{T}(t)\rho =\exp [t\mathcal{D}]\rho .  \label{UnormalizadaT}
\end{equation}%
Here, the exponential superoperator is defined by the generator $\mathcal{D}%
, $ which is the complement of $\mathcal{J},$ that is, $\mathcal{L}=\mathcal{%
D}+\mathcal{J}.$ Hence, from Eq. (\ref{LindbladInteraction}) it reads%
\begin{equation}
\mathcal{D}\rho =(\mathcal{L}_{s}+\mathcal{L}_{a})\rho -\frac{\gamma }{2}%
\{\left\vert a_{\mathrm{0}}\right\rangle \left\langle a_{\mathrm{0}%
}\right\vert ,\rho \}_{+}.  \label{DBipartito}
\end{equation}

The measurement transformation $\mathcal{M}$\ and the propagator $\mathcal{T}%
_{c}(t)$\ completely define the structure of the realizations of $\rho
_{sa}^{\mathrm{st}}(t).$\ It only remains to define the algorithm that
allows to obtain the random detection times. Here they are characterized
through a survival probability function $P_{0}(t|\rho )$ \cite{plenio}.
Given that at time $\tau $ the bipartite system state is $\rho ,$ the
probability of not happening any detection up to time $t$ is \cite%
{carmichael}%
\begin{equation}
P_{0}(t-\tau |\rho )=\mathrm{Tr}_{sa}[\mathcal{T}(t-\tau )\rho ]=\mathrm{Tr}%
_{sa}[e^{t\mathcal{D}}\rho ].  \label{SurvivalBipartita}
\end{equation}%
With this function the realizations can be obtained as follows. Given the
initial state $\rho _{0}^{sa},$ the time $t_{1}$ of the first detection
event follows by solving the equation $P_{0}(t_{1}-0|\rho _{0}^{sa})=r,$
where $r$ is a random number in the interval $(0,1).$ The dynamic of $\rho
_{sa}^{\mathrm{st}}(t)$\ in the interval $(0,t_{1})$ is defined by Eq. (\ref%
{Conditional}). At $t=t_{1}$ the disruptive transformation [Eq. (\ref%
{MBpartito})] $\rho _{s}^{\mathrm{st}}(t_{1})\rightarrow \mathcal{M}\rho
_{sa}^{\mathrm{st}}(t_{1})$ is applied. The subsequent dynamics is the same.
In fact, after the $n_{th}-$measurement event at time $t_{n},$ $\rho _{sa}^{%
\mathrm{st}}(t_{n})\rightarrow \mathcal{M}\rho _{sa}^{\mathrm{st}}(t_{n}),$
the time $t_{n+1}$ for the next detection event follows from $%
P_{0}(t_{n+1}-t_{n}|\mathcal{M}\rho _{sa}^{\mathrm{st}}(t_{n}))=r,$ where
again $r$ is a random number in the interval $(0,1).$ The dynamic in the
interval $(t_{n},t_{n+1})$ is defined by the conditional propagator (\ref%
{Conditional}). The realizations generated with this algorithm fulfill Eq. (%
\ref{RhoAverage}) (see for example Appendix A of Ref. \cite{OneChannel}).

\subsection{Stochastic realizations}

The standard quantum jump approach allows to defining the realizations of $%
\rho _{sa}^{\mathrm{st}}(t).$ Straightforwardly from this object it is
possible to obtain the partial stochastic dynamics of each system,%
\begin{equation}
\rho _{s}^{\mathrm{st}}(t)=\mathrm{Tr}_{a}[\rho _{sa}^{\mathrm{st}}(t)],\ \
\ \ \ \ \ \ \rho _{a}^{\mathrm{st}}(t)=\mathrm{Tr}_{s}[\rho _{sa}^{\mathrm{st%
}}(t)].
\end{equation}%
Furthermore, from Eq. (\ref{RhoAverage}), the relations $\rho _{t}^{s}=%
\overline{\rho _{s}^{\mathrm{st}}(t)},$ and $\rho _{t}^{a}=\overline{\rho
_{a}^{\mathrm{st}}(t)}$ are also valid. Given the separable initial
condition (\ref{CISeparable}), from Eqs. (\ref{MBpartito}) and (\ref%
{DBipartito}) it is simple to realize that $\rho _{sa}^{\mathrm{st}}(t)$
becomes separable at all times%
\begin{equation}
\rho _{sa}^{\mathrm{st}}(t)=\rho _{s}^{\mathrm{st}}(t)\otimes \rho _{a}^{%
\mathrm{st}}(t).  \label{Separability}
\end{equation}%
In fact, given the absence of initial correlations, the conditional dynamic (%
\ref{Conditional}) remains separable [see Eq. (\ref{DBipartito})].
Furthermore, in each detection event, given a separable input, the post
measurement state also becomes separable. Nevertheless, notice that $\rho
_{s}^{\mathrm{st}}(t)$ and $\rho _{a}^{\mathrm{st}}(t)$ are statistically
correlated. Below, we describe their dynamics.

\subsubsection{Ancilla realizations}

After taking a partial trace over Eq. (\ref{MBpartito}), from Eq. (\ref%
{Separability}) we deduce that in each measurement event the ancilla state
suffer the transformation%
\begin{equation}
\rho _{a}^{\mathrm{st}}(t)\rightarrow \mathrm{Tr}_{s}[\mathcal{M}\rho _{sa}^{%
\mathrm{st}}(t)]=\frac{\mathcal{J}_{a}\rho _{a}^{\mathrm{st}}(t)}{\mathrm{Tr}%
_{a}[\mathcal{J}_{a}\rho _{a}^{\mathrm{st}}(t)]}=\bar{\rho}_{a},
\label{AncillaTransformation}
\end{equation}%
where the ancilla superoperator $\mathcal{J}_{a}$ is%
\begin{equation}
\mathcal{J}_{a}[\rho ]=\gamma \left\langle a_{\mathrm{0}}\right\vert \rho
\left\vert a_{\mathrm{0}}\right\rangle \bar{\rho}_{a}.  \label{JAncilla}
\end{equation}%
Hence, the collapsed ancilla state is always the same [Eq. (\ref%
{AncillaReset})]. Similarly, from Eqs. (\ref{UnormalizadaT}) and (\ref%
{DBipartito}), we deduce that between detection events the conditional
ancilla dynamics is defined by the (unnormalized) superoperator $\mathcal{T}%
_{a}(t)\rho =\exp [t\mathcal{D}_{a}]\rho ,$ where%
\begin{equation}
\mathcal{D}_{a}\rho =\mathcal{L}_{a}\rho -\frac{\gamma }{2}\{\left\vert a_{%
\mathrm{0}}\right\rangle \left\langle a_{\mathrm{0}}\right\vert ,\rho \}_{+}.
\label{DAncilla}
\end{equation}%
For separable initial conditions, this propagator also applies at the
initial time. This simplification explain the chosen initial state (\ref%
{CISeparable}) and the projectors (\ref{projectors}).

From Eqs. (\ref{DBipartito}) and (\ref{SurvivalBipartita}), we notice that
the survival probability can be rewritten as $[P_{0}(t-\tau |\rho
)\rightarrow P_{0}(t-\tau )]$%
\begin{equation}
P_{0}(t-\tau )=\mathrm{Tr}_{a}[\exp [\mathcal{D}_{a}(t-\tau )]\bar{\rho}%
_{a}].  \label{SurvivalCYQRW}
\end{equation}%
In fact, the ancilla state is always the same after a detection event.
Consequently, the measurement statistics correspond to a renewal process,
that is, the inter-event probability distribution is always the same. On the
other hand, it is simple to realize that the measurement transformation (\ref%
{AncillaTransformation}), the conditional ancilla dynamics defined by Eq. (%
\ref{DAncilla}), and the survival probability (\ref{SurvivalCYQRW}) also
arise by formulating the quantum jump approach over the basis of Eq. (\ref%
{AncillaEvolution}). In fact, $\mathbb{L}_{a}=\mathcal{D}_{a}+\mathcal{J}%
_{a}.$

\subsubsection{System realizations}

Given the separability property (\ref{Separability}), from Eq. (\ref%
{MBpartito}) it follows that in each detection event (ancilla measurement
apparatus), the system suffer the transformation%
\begin{equation}
\rho _{s}^{\mathrm{st}}(t)\rightarrow \mathrm{Tr}_{a}[\mathcal{M}\rho _{sa}^{%
\mathrm{st}}(t)]=\mathcal{E}_{s}[\rho _{s}^{\mathrm{st}}(t)],
\label{SystemTransformation}
\end{equation}%
that is, the transformation associated to a collision event. On the other
hand, given that a measurement event happened at time $\tau ,$ from Eqs. (%
\ref{Conditional}) and (\ref{DBipartito}) we deduce that the posterior
system conditional evolution is given by%
\begin{equation}
\rho _{s}^{\mathrm{st}}(t)=\mathrm{Tr}_{a}[\mathcal{T}_{c}(t-\tau )\rho
_{sa}^{\mathrm{st}}(\tau )]=\exp [\mathcal{L}_{s}(t-\tau )]\rho _{s}^{%
\mathrm{st}}(\tau ).  \label{IntereventSystemDynamics}
\end{equation}%
This inter-event evolution also correspond to the dynamics of the QCM.
Therefore, by assuming that the measurement process is performed over the
ancilla system, the realizations of the system of interest have the same
structure than in the phenomenological QCM. This is the main result of this
section. Notice that each system collisional event happens when the
measurement apparatus detects an ancilla transition.

The renewal property of the realizations was proven previously. In fact,
form the survival probability (\ref{SurvivalCYQRW}) we define the waiting
time distribution $w(t)=-(d/dt)P_{0}(t),$ which delivers 
\begin{subequations}
\label{waiting}
\begin{eqnarray}
w(t) &=&-\mathrm{Tr}_{a}[\mathcal{D}_{a}\exp [t\mathcal{D}_{a}]\bar{\rho}%
_{a}], \\
&=&\gamma \left\langle a_{\mathrm{0}}\right\vert \exp (t\mathcal{D}_{a})[%
\bar{\rho}_{a}]\left\vert a_{\mathrm{0}}\right\rangle .
\end{eqnarray}%
In deriving this expression we used Eq. (\ref{DAncilla}) and that $\mathrm{Tr%
}_{a}[\mathcal{L}_{a}\rho ]=0.$ Hence, in the present modeling the quantum
jump approach allows to write the waiting time distribution in terms of the
ancilla dynamics. Indeed, from Eqs. (\ref{SystemTransformation}) and (\ref%
{IntereventSystemDynamics}), we deduce that the ancilla dynamics mainly
determine the statistic of the system realizations.

\subsection{Consistence between master equation and ensemble of realizations}

For showing the consistence of the developed results, it remains to
demonstrate that the waiting time distribution (\ref{waiting}), which
determine the realizations statistics, and the kernel (\ref{LindbladKernel}%
), which determine the density matrix evolution, fulfill in the Laplace
domain the relation (\ref{CollisionSystemWaiting}).

The Laplace transform of Eq. (\ref{waiting}) reads 
\end{subequations}
\begin{equation}
w(u)=\gamma \left\langle a_{\mathrm{0}}\right\vert \frac{1}{u-\mathcal{D}_{a}%
}[\bar{\rho}_{a}]\left\vert a_{\mathrm{0}}\right\rangle ,
\end{equation}%
while from Eq. (\ref{LindbladKernel}) we obtain%
\begin{equation}
\frac{k(u)}{u}=\gamma \left\langle a_{\mathrm{0}}\right\vert \frac{1}{u-%
\mathbb{L}_{a}}[\bar{\rho}_{a}]\left\vert a_{\mathrm{0}}\right\rangle .
\label{KernelLaplacero}
\end{equation}%
In deriving this expression we used that $\left\langle a_{\mathrm{0}%
}\right\vert \bar{\rho}_{a}\left\vert a_{\mathrm{0}}\right\rangle =0.$ On
the other hand, using that $\mathbb{L}_{a}=\mathcal{D}_{a}+\mathcal{J}_{a}$
it follows the relation%
\begin{equation}
\frac{1}{u-\mathbb{L}_{a}}=\sum_{n=0}^{\infty }\Big{[}\frac{1}{u-\mathcal{D}%
_{a}}\mathcal{J}_{a}\Big{]}^{n}\frac{1}{u-\mathcal{D}_{a}}.
\end{equation}%
By introducing this expression in Eq. (\ref{KernelLaplacero}) and by using
the definition (\ref{JAncilla}) we get%
\begin{equation}
\frac{k(u)}{u}=\sum_{n=1}^{\infty }w(u)=\frac{w(u)}{1-w(u)},
\end{equation}%
which recovers the relation (\ref{CollisionSystemWaiting}) associated to the
phenomenological approach.

\section{Example}

In this section, we study the dynamics of a two-level system, which in turn
may be read, for example, as a qubit unit. In quantum information
arrangements it is expected that decoherence and dissipation are
\textquotedblleft mediated\textquotedblright\ by interactions with extra
quantum subunits. Therefore, as ancilla we consider another system whose
dynamics is able to develops quantum coherent effects. For simplicity it is
also taken as a two-level system.

In the approach developed in the previous sections, the collision statistics
is completely defined by the ancilla dynamics. Hence, in the next example,
it structure depends on underlying quantum coherent effects. We remark that
this feature is foreign in phenomenological formulations where the waiting
time distribution is usually defined by a linear combination of exponential
functions \cite{classBu,GrigoBu,VacCol,giraldi}. We demonstrate that this
kind of statistics arise when the ancilla dynamics is completely incoherent.
This property motivate the dynamics studied below. Both dephasing and
dissipative channels are formulated.

\subsection{Dephasing channel}

As system we consider a two-level system whose Hamiltonian reads $%
H_{s}=\hbar \omega _{s}\sigma _{z}/2,$ where $\omega _{s}$ is the transition
frequency between its eigenstates, denoted as $\left\vert \pm \right\rangle
, $ while $\sigma _{z}$\ is the $z$-Pauli matrix. The ancilla system is also
a two-level system. In an interaction representation with respect to $H_{s}$
the evolution of the bipartite state $\rho _{t}^{sa}$ reads%
\begin{equation}
\frac{d\rho _{t}^{sa}}{dt}=\frac{-i\Delta }{2}[\mathrm{I}_{s}\otimes \sigma
_{x},\rho _{t}^{sa}]+\gamma ([V,\rho _{t}^{sa}V^{\dagger }]+[V\rho
_{t}^{sa},V^{\dagger }]).  \label{LindbladEjemplo}
\end{equation}%
The first unitary contribution defines the ancilla Hamiltonian. It is given
by the $x$-Pauli matrix $\sigma _{x}$ written in the basis of $\sigma _{z}$
eigenstates: $\left\vert \pm \right\rangle .$ The Lindblad contribution is
written in terms of the operator [see Eq. (\ref{Operators})]%
\begin{equation}
V=\sigma _{z}\otimes \sigma .  \label{V_Bipartito}
\end{equation}%
Here, $\sigma =\left\vert -\right\rangle \left\langle +\right\vert $ is the
lowering operator acting on the ancilla states $\left\vert \pm \right\rangle
.$ Hence, $V$ leads to a dissipative coupling between both systems. The
initial bipartite state [see Eq. (\ref{CISeparable})] is taken as%
\begin{equation}
\rho _{0}^{sa}=\rho _{0}^{s}\otimes \left\vert -\right\rangle \left\langle
-\right\vert ,  \label{intialEstocastico}
\end{equation}%
where $\rho _{0}^{s}$\ is an arbitrary system state. The ancilla begins in
its lower state.

Performing the partial trace $\rho _{t}^{a}=\mathrm{Tr}_{s}\left[ \rho
_{t}^{sa}\right] ,$ the bipartite evolution (\ref{LindbladEjemplo}) leads to%
\begin{equation}
\frac{d\rho _{t}^{a}}{dt}=\frac{-i\Delta }{2}[\sigma _{x},\rho
_{t}^{a}]+\gamma ([\sigma ,\rho _{t}^{a}\sigma ^{\dagger }]+[\sigma \rho
_{t}^{a},\sigma ^{\dagger }]).  \label{AncillaFluor}
\end{equation}%
This marginal ancilla dynamics corresponds to a quantum fluorescent system 
\cite{breuerbook,carmichaelbook}, where $\gamma $ defines its natural decay
rate while $\Delta $ is the Rabi frequency. On the other hand, the
interaction defined by Eq. (\ref{V_Bipartito}) lead to a decoherence system
channel \cite{buzeMarkov}. Hence, only the system coherences are affected by
the undesirable interaction.

\subsubsection{System stochastic realizations}

The measurement apparatus record the ancilla transitions $\left\vert
+\right\rangle \rightsquigarrow \left\vert -\right\rangle .$ Therefore, from
Eqs. (\ref{MBpartito}) and (\ref{V_Bipartito}) we deduce that in each
measurement event the ancilla collapse to its ground state $\bar{\rho}%
_{a}=\left\vert -\right\rangle \left\langle -\right\vert ,$ while the system
suffer the completely positive transformation%
\begin{equation}
\mathcal{E}_{s}[\rho ]=\sigma _{z}\rho \sigma _{z}.  \label{SuperZetal}
\end{equation}%
As is well known, this superoperator lead to a change of sign in the system
coherences \cite{classBu}. On the other hand, during the successive
measurement events the system dynamics is frozen, that is, it does not
evolves. This conclusion follows from Eq. (\ref{IntereventSystemDynamics})
and (\ref{LindbladEjemplo}).

The statistics of the time interval between successive detections events
define a renewal process. Its probability distribution is given by Eq. (\ref%
{waiting}). Under the associations $\left\vert a_{\mathrm{0}}\right\rangle
\rightarrow \left\vert +\right\rangle ,$ and $\mathcal{D}_{a}[\rho
]=-(i\Delta /2)[\sigma _{x},\rho ]-(1/2)\gamma \{\sigma ^{\dagger }\sigma
,\rho \}_{+},$ we get the waiting time distribution%
\begin{equation}
w(t)=4\gamma \Delta ^{2}e^{-\gamma t/2}\left\{ \frac{\sinh [(t/4)\sqrt{%
\gamma ^{2}-4\Delta ^{2}}]}{\sqrt{\gamma ^{2}-4\Delta ^{2}}}\right\} ^{2}.
\label{WaiterFluor}
\end{equation}

Notice that Eqs. (\ref{SuperZetal}) and (\ref{WaiterFluor}) completely
define the system realizations.%
%figura%figura%figura%figura%figurav%figura%figura%figura%figura%figura%figura%figura%figura%figura%figurav%figura%figura%figura%figura%figura
%figura%figura%figura%figura%figurav%figura%figura%figura%figura%figura%figura%figura%figura%figura%figurav%figura%figura%figura%figura%figura
\begin{figure}[tbp]
\includegraphics[bb=28 146 371 540,width=8.5cm]{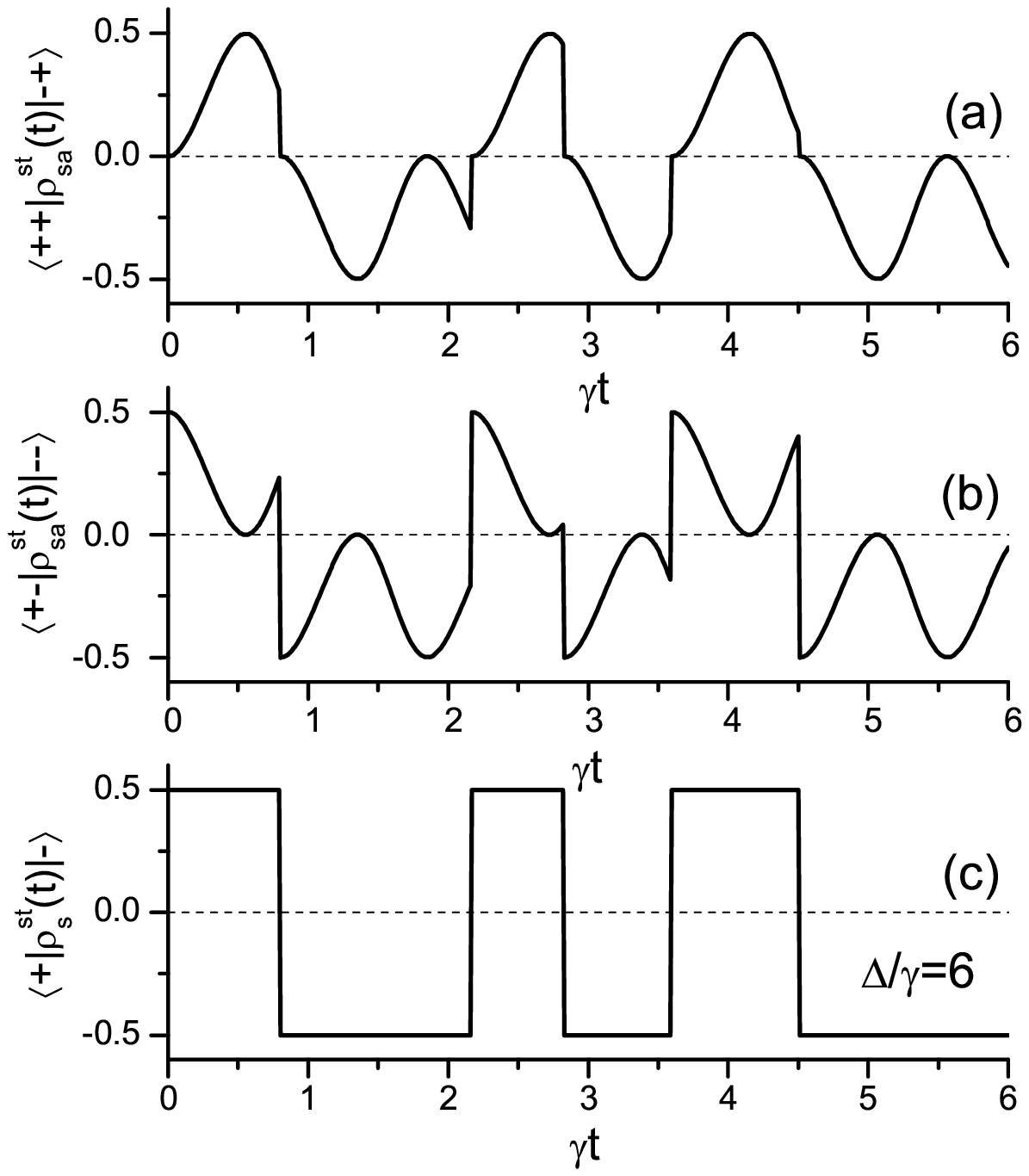}
\caption{Realizations of matrix elements of the stochastic density matrix $%
\protect\rho _{\mathrm{st}}^{sa}(t)$ and $\protect\rho _{\mathrm{st}%
}^{s}(t). $ (a) $\left\langle ++\right\vert \protect\rho _{sa}^{\mathrm{st}%
}(t)\left\vert -+\right\rangle .$ (b) $\left\langle +-\right\vert \protect%
\rho _{sa}^{\mathrm{st}}(t)\left\vert --\right\rangle .$\ (c) $\left\langle
+\right\vert \protect\rho _{s}^{\mathrm{st}}(t)\left\vert -\right\rangle .$
The characteristic parameters of the bipartite evolution (\protect\ref%
{LindbladEjemplo}) satisfy $\Delta /\protect\gamma =6.$}
\end{figure}
%figura%figura%figura%figura%figurav%figura%figura%figura%figura%figura%figura%figura%figura%figura%figurav%figura%figura%figura%figura%figura
%figura%figura%figura%figura%figurav%figura%figura%figura%figura%figura%figura%figura%figura%figura%figurav%figura%figura%figura%figura%figura
%figura%figura%figura%figura%figurav%figura%figura%figura%figura%figura%figura%figura%figura%figura%figurav%figura%figura%figura%figura%figura
%figura%figura%figura%figura%figurav%figura%figura%figura%figura%figura%figura%figura%figura%figura%figurav%figura%figura%figura%figura%figura

In Fig. 1 we show a realization of the system coherence $\left\langle
+\right\vert \rho _{s}^{\mathrm{st}}(t)\left\vert -\right\rangle .$ In order
to show the consistence of the developed approach, it was obtained from the
realizations of the underlying bipartite dynamics,%
\begin{equation}
\left\langle +\right\vert \rho _{s}^{\mathrm{st}}(t)\left\vert
-\right\rangle =\left\langle ++\right\vert \rho _{sa}^{\mathrm{st}%
}(t)\left\vert -+\right\rangle +\left\langle +-\right\vert \rho _{sa}^{%
\mathrm{st}}(t)\left\vert --\right\rangle ,  \label{MarginalCoherencia}
\end{equation}%
that is, from the partial trace of $\rho _{sa}^{\mathrm{st}}(t).$ The states 
$\{\left\vert ij\right\rangle \},$ $i,j=+,-,$ provide a complete basis of
the bipartite Hilbert space. The realizations of $\rho _{sa}^{\mathrm{st}%
}(t) $ follows from a \textquotedblleft standard Markovian quantum jump
approach\textquotedblright\ formulated on the basis of Eq. (\ref%
{LindbladEjemplo}). We have taken the initial condition $\rho _{sa}^{\mathrm{%
st}}(0)=\left\vert x_{+}\right\rangle \left\langle x_{+}\right\vert \otimes
\left\vert -\right\rangle \left\langle -\right\vert ,$ where $\left\vert
x_{+}\right\rangle =(1/\sqrt{2})(\left\vert +\right\rangle +\left\vert
-\right\rangle )$ is an eigenstate of $\sigma _{x}.$ In Fig. 1(a), we see
that in each recording event the bipartite coherence $\left\langle
++\right\vert \rho _{sa}^{\mathrm{st}}(t)\left\vert -+\right\rangle $
collapse to zero,%
\begin{equation}
\left\langle ++\right\vert \mathcal{M}\rho _{sa}^{\mathrm{st}}(t)\left\vert
-+\right\rangle =0.
\end{equation}%
This result follows from the action of the operator (\ref{V_Bipartito}),
which induces the ancilla transitions $\left\vert +\right\rangle
\rightsquigarrow \left\vert -\right\rangle .$\ On the other, the bipartite
coherence $\left\langle +-\right\vert \rho _{sa}^{\mathrm{st}}(t)\left\vert
--\right\rangle $ suffers the disruptive changes $\left\langle +-\right\vert
\rho _{sa}^{\mathrm{st}}(t)\left\vert --\right\rangle \rightarrow
-\left\langle +-\right\vert \rho _{sa}^{\mathrm{st}}(0)\left\vert
--\right\rangle ,$ Fig. 1(b). By calculating the measurement transformation (%
\ref{MBpartito}), from Eq. (\ref{V_Bipartito}) we get%
\begin{equation*}
\left\langle +-\right\vert \mathcal{M}\rho _{sa}^{\mathrm{st}}\left\vert
--\right\rangle =\frac{-\left\langle ++\right\vert \rho _{sa}^{\mathrm{st}%
}\left\vert -+\right\rangle }{\left\langle ++\right\vert \rho _{sa}^{\mathrm{%
st}}\left\vert ++\right\rangle +\left\langle -+\right\vert \rho _{sa}^{%
\mathrm{st}}\left\vert -+\right\rangle }.
\end{equation*}%
By an explicitly calculation of the conditional evolution defined by the
operator $\mathcal{D},$ Eq. (\ref{DBipartito}), if follows that the quotient
of the previous bipartite matrix elements is an invariant of the conditional
evolution, delivering the observed property%
\begin{equation}
\left\langle +-\right\vert \mathcal{M}\rho _{sa}^{\mathrm{st}}(t)\left\vert
--\right\rangle =-\left\langle +\right\vert \rho _{s}^{\mathrm{st}%
}(0)\left\vert -\right\rangle ,
\end{equation}%
where we have used that $\left\langle +-\right\vert \rho _{sa}^{\mathrm{st}%
}(0)\left\vert --\right\rangle =\left\langle +\right\vert \rho _{s}^{\mathrm{%
st}}(0)\left\vert -\right\rangle $ [Eq. (\ref{intialEstocastico})].
Therefore, in each measurement event the coherence $\left\langle
+-\right\vert \rho _{sa}^{\mathrm{st}}(t)\left\vert --\right\rangle ,$
beside a change of sign, recovers its initial value.

In Fig. 1(c), we plot the realization of $\left\langle +\right\vert \rho
_{s}^{\mathrm{st}}(t)\left\vert -\right\rangle $ obtained from Eq. (\ref%
{MarginalCoherencia}), that is by adding the two bipartite coherences. As
both coherences $\left\langle ++\right\vert \mathcal{M}\rho _{sa}^{\mathrm{st%
}}(t)\left\vert -+\right\rangle $ and $\left\langle +-\right\vert \mathcal{M}%
\rho _{sa}^{\mathrm{st}}(t)\left\vert --\right\rangle $\ always oscillate in
a complementary way, during the inter-event time intervals $\left\langle
+\right\vert \rho _{s}^{\mathrm{st}}(t)\left\vert -\right\rangle $ is
constant, while in the measurement events it changes of sign. In this way,
we explicitly show that the underlying quantum jump approach lead to the
realizations of the phenomenological collision model. In fact, the action of
the superoperator (\ref{SuperZetal}) only introduce a change of sign in the
system coherences. In a similar way, it is possible to show that the system
populations are not affected by the dynamics, that is, $\left\langle \pm
\right\vert \rho _{s}^{\mathrm{st}}(t)\left\vert \pm \right\rangle
=\left\langle \pm \right\vert \rho _{s}^{\mathrm{st}}(0)\left\vert \pm
\right\rangle .$

\subsubsection{Density matrix evolution}

In Fig. 2 we show the average coherence behavior obtained from the ensemble
of realizations shown in Fig. 1 (noisy curve). Furthermore, we present the
exact solution of the coherence that follows from the master equation (\ref%
{MasterColision}) (black full line). Taking into account the underlying
Lindblad equation (\ref{LindbladEjemplo}), it can be written as%
\begin{equation}
\frac{d}{dt}\rho _{t}^{s}=\int_{0}^{t}dt^{\prime }k(t-t^{\prime })\mathcal{C}%
_{s}[\rho _{t^{\prime }}^{s}].  \label{DispersivaNoMarkoviana}
\end{equation}%
The superoperator $\mathcal{C}_{s}=(\mathcal{E}_{s}-\mathrm{I}_{s}),$ from
Eq. (\ref{SuperZetal}) reads%
\begin{equation}
\mathcal{C}_{s}[\bullet ]=\frac{1}{2}([\sigma _{z},\bullet \sigma
_{z}]+[\sigma _{z}\bullet ,\sigma _{z}]).  \label{CDispersivo}
\end{equation}%
On the other hand, the kernel is determined by the general expression (\ref%
{LindbladKernel}). From Eq. (\ref{AncillaFluor}) it follows%
\begin{equation}
k(t)=2\gamma \Delta ^{2}e^{-(3/4)\gamma t}\left\{ \frac{\sinh [(t/4)\sqrt{%
\gamma ^{2}-16\Delta ^{2}}]}{\sqrt{\gamma ^{2}-16\Delta ^{2}}}\right\} .
\end{equation}%
This kernel and the waiting time distribution (\ref{WaiterFluor}) fulfill
the Laplace relation (\ref{CollisionSystemWaiting}). 
%figura%figura%figura%figura%figurav%figura%figura%figura%figura%figura%figura%figura%figura%figura%figurav%figura%figura%figura%figura%figura
%figura%figura%figura%figura%figurav%figura%figura%figura%figura%figura%figura%figura%figura%figura%figurav%figura%figura%figura%figura%figura
\begin{figure}[tbp]
\includegraphics[bb=16 188 426 541,width=8.5cm]{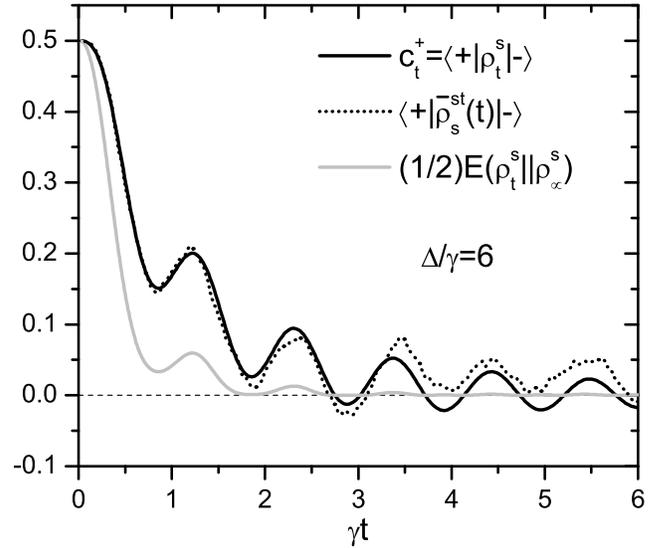}
\caption{System coherence. Full line, exact solution Eq. (\protect\ref%
{CoherSolution}). Dotted (noisy) line, average coherence $\left\langle
+\right\vert \overline{\protect\rho _{s}^{\mathrm{st}}(t)}\left\vert
-\right\rangle $ obtained by averaging $10^{3}$ realizations. Grey line,
relative entropy $E(\protect\rho _{t}^{s}||\protect\rho _{\infty }^{s}),$
Eq. (\protect\ref{Relativa}). The parameters are the same than in Fig. 1, $%
\Delta /\protect\gamma =6.$}
\end{figure}
%figura%figura%figura%figura%figurav%figura%figura%figura%figura%figura%figura%figura%figura%figura%figurav%figura%figura%figura%figura%figura
%figura%figura%figura%figura%figurav%figura%figura%figura%figura%figura%figura%figura%figura%figura%figurav%figura%figura%figura%figura%figura
%figura%figura%figura%figura%figurav%figura%figura%figura%figura%figura%figura%figura%figura%figura%figurav%figura%figura%figura%figura%figura
%figura%figura%figura%figura%figurav%figura%figura%figura%figura%figura%figura%figura%figura%figura%figurav%figura%figura%figura%figura%figura

Consistently with the system stochastic realizations, Eq. (\ref%
{DispersivaNoMarkoviana}) does not modify the populations, $\left\langle \pm
\right\vert \rho _{t}^{s}\left\vert \pm \right\rangle =\left\langle \pm
\right\vert \rho _{0}^{s}\left\vert \pm \right\rangle .$ On the other hand,
working in a Laplace domain, the coherences $c_{t}^{\pm }\equiv \left\langle
\pm \right\vert \rho _{t}^{s}\left\vert \mp \right\rangle $ read%
\begin{eqnarray}
c_{t}^{\pm } &=&c_{0}^{\pm }\left\{ e^{-\gamma t}\frac{2\Delta ^{2}}{\gamma
^{2}+2\Delta ^{2}}+e^{-\gamma t/4}\right.  \label{CoherSolution} \\
&&\!\left. \times \Big{[}\frac{\gamma ^{2}}{\gamma ^{2}+2\Delta ^{2}}\cosh
(\varphi t)+\frac{\gamma (\gamma ^{2}+8\Delta ^{2})}{4(\gamma ^{2}+2\Delta
^{2})}\frac{\sinh (\varphi t)}{\varphi }\Big{]}\!\right\} ,  \notag
\end{eqnarray}%
where for shortening the expression we introduced the \textquotedblleft
frequency\textquotedblright 
\begin{equation}
\varphi =\sqrt{(\gamma /4)^{2}-\Delta ^{2}}.  \label{phi}
\end{equation}%
The same expression follows from the alternative solution $c_{t}^{\pm
}=\left\langle \pm \right\vert \rho _{t}^{s}\left\vert \mp \right\rangle
=\left\langle \pm \pm \right\vert \rho _{t}^{sa}\left\vert \mp \pm
\right\rangle +\left\langle \pm \mp \right\vert \rho _{t}^{sa}\left\vert \mp
\mp \right\rangle ,$ where $\rho _{t}^{sa}$\ is the solution of the
bipartite evolution (\ref{LindbladEjemplo}). Notice that in Eq. (\ref%
{CoherSolution}), besides a monotonic decaying contribution, the two
remaining terms may develop an oscillatory behavior. As shown in Fig. 2, Eq.
(\ref{CoherSolution}) correctly fits the average ensemble behavior.

\subsubsection{Environment-to-system back flow of information}

The analysis of Refs. \cite{pheno,pheno1} demonstrate that QCMs may lead to
non-Markovian effects such as an environment-to-system back flow of
information \cite{NoMeasure}. This property or phenomenon can be defined on
the basis of \textquotedblleft any measure\textquotedblright\ that in the
Markovian case present a monotonic time decay behavior \cite{breuerbook}.
One well known example is the relative entropy between two states \cite%
{Entropy,pheno}. As we are not interested in quantifying the non-Markovian
effects, for simplicity here we consider the relative entropy with respect
to the stationary state%
\begin{equation}
E(\rho _{t}^{s}||\rho _{\infty }^{s})=\mathrm{Tr}_{s}[\rho _{t}^{s}(\ln
_{2}\rho _{t}^{s}-\ln _{2}\rho _{s}^{\infty })],  \label{Relativa}
\end{equation}%
where $\rho _{s}^{\infty }=\lim_{t\rightarrow \infty }\rho _{t}^{s}.$ Hence,
the back flow of information arises if there exists times $t_{2}>t_{1}$ such
that $E(\rho _{t_{2}}^{s}||\rho _{\infty }^{s})>E(\rho _{t_{1}}^{s}||\rho
_{\infty }^{s}).$ Below we show that this feature arises in the dynamics
described previously.

In Fig. 2 we also plotted $E(\rho _{t}^{s}||\rho _{\infty }^{s})$\ (grey
full line) where $\rho _{t}^{s}$ is the solution of the Eq. (\ref%
{DispersivaNoMarkoviana}). The stationary state is the diagonal matrix $\rho
_{\infty }^{s}=\mathrm{diag}\{\left\langle +\right\vert \rho
_{0}^{s}\left\vert +\right\rangle ,\left\langle -\right\vert \rho
_{0}^{s}\left\vert -\right\rangle \}.$ Clearly the time behavior is
non-monotonous, indicating a back-flow of information. Furthermore, the
oscillatory behavior of $E(\rho _{t}^{s}||\rho _{\infty }^{s})$ is
correlated with the oscillatory behavior of the coherences, which arise\
whenever $\varphi $ is a complex quantity, that is, from Eq. (\ref{phi}), $%
\Delta >(\gamma /4).$

%The oscillatory behavior of $E(\rho _{t_{2}}^{s}||\rho _{\infty }^{s})$ does
%not relies in the dephasing character of the evolution. In fact, by adding
%to Eq. (\ref{LindbladEjemplo}) the system Hamiltonian $H_{s}=\hbar \Omega
%\sigma _{x}/2,$ the dynamic losses it dephasing character. The stationary
%populations and coherences only depend on the characteristic parameters of
%the problem. Furthermore, Eq. (\ref{DispersivaNoMarkoviana}) acquires the
%structure (\ref{MasterColision}). Indeed, the only change is the
%inter-collision dynamics, which is given by Eq. (\ref%
%{IntereventSystemDynamics}), that is, $\exp [-it\Omega \sigma _{x}/2]\bullet
%\exp [+it\Omega \sigma _{x}/2].$ In this situation, we also find that $%
%E(\rho _{t}^{s}||\rho _{\infty }^{s})$ does not decay in a monotonous way,
%showing that in general, the master equations associated to the QCM may lead
%to a back-flow of information from the system to the environment \cite%
%{pheno,pheno1}.

\subsubsection{Incoherent ancilla dynamics}

For the dynamics (\ref{LindbladEjemplo}), the ancilla dynamics develops
quantum coherent effects, Eq. (\ref{AncillaFluor}), which in turn determine
the waiting time distribution, Eq. (\ref{WaiterFluor}). Here, we introduce
an alternative ancilla dynamics which only induces incoherent transitions.
Instead of Eq. (\ref{LindbladEjemplo}), for the same system $S,$ we take the
bipartite evolution as%
\begin{eqnarray}
\frac{d\rho _{t}^{sa}}{dt} &=&\gamma ([V,\rho _{t}^{sa}V^{\dagger }]+[V\rho
_{t}^{sa},V^{\dagger }]) \\
&&+\beta ([A,\rho _{t}^{sa}A^{\dagger }]+[A\rho _{t}^{sa},A^{\dagger }]), 
\notag
\end{eqnarray}%
with initial condition $\rho _{0}^{sa}=\rho _{0}^{s}\otimes \left\vert
-\right\rangle \left\langle -\right\vert ,$ while%
\begin{equation}
V=\sigma _{z}\otimes \sigma ,\ \ \ \ \ \ \ \ \ \ \ A=\mathrm{I}_{s}\otimes
\sigma ^{\dagger }.  \label{opereta}
\end{equation}%
Hence, the ancilla dynamics [Eq. (\ref{AncillaEvolution})] only leads to the
incoherent (classical) transitions $\left\vert +\right\rangle \overset{%
\gamma }{\rightsquigarrow }\left\vert -\right\rangle $ and $\left\vert
-\right\rangle \overset{\beta }{\rightsquigarrow }\left\vert +\right\rangle
. $ Its statistical behavior is defined by a (two-level) classical rate
master equation.

We assume that the recording apparatus is only sensitive to the ancilla
transition $\left\vert +\right\rangle \rightsquigarrow \left\vert
-\right\rangle ,$ that is, the transition induced by the operator $V.$ In
this situation, from Eqs. (\ref{MBpartito}) and (\ref{opereta}), we deduce
that the collisional superoperator again reads $\mathcal{E}_{s}[\rho
]=\sigma _{z}\rho \sigma _{z}$ [Eq. (\ref{SuperZetal})]. Thus, the system
evolution is given by Eq. (\ref{DispersivaNoMarkoviana}). Nevertheless, the
kernel follows from Eq. (\ref{CollisionSystemWaiting}), where the waiting
time distribution can be calculated from Eq. (\ref{waiting}). We get%
\begin{equation}
w(u)=\left( \frac{\gamma }{u+\gamma }\right) \left( \frac{\beta }{u+\beta }%
\right) .  \label{WaiterClasica}
\end{equation}%
In the time domain $w(t)$ is the convolution of two exponential functions.
The system coherences become $c_{u}^{\pm }=c_{0}^{\pm }(u+\gamma +\beta
)/[2u^{2}+2u(\gamma +\beta )+\gamma \beta ],$ which can be written as a
lineal combination of exponential functions. Independently of the initial
conditions, in this case the dynamics does not present an
environment-to-system back flow of information, suggesting that underlying
coherent effects may be necessary for the development of this phenomenon.

Taking an ancilla system with higher number of states, all of them coupled
via incoherent transitions, the waiting time distribution results defined by
more complex expressions which in the time domain are linear combinations of
exponential functions. For example, taking an unidirectional coupling $%
\left\vert a_{0}\right\rangle \rightsquigarrow \left\vert a_{1}\right\rangle
\rightsquigarrow \cdots \left\vert a_{m}\right\rangle \rightsquigarrow
\left\vert a_{0}\right\rangle ,$ all of them with rate $\gamma ,$ the
waiting time distribution becomes $w(u)=[\gamma /(u+\gamma )]^{m+1}.$ This
kind of distributions, which rely on incoherent ancilla dynamics, were
considered, for example, in Ref. \cite{VacCol}.

\subsection{Dissipative channels}

In the previous example, Eqs. (\ref{DispersivaNoMarkoviana}) and (\ref%
{CDispersivo}) define a non-Markovian decoherence channel. One may also
consider interactions that lead to dissipative channels. For example,
maintaining the ancilla dynamics (\ref{AncillaFluor}), a depolarizing \cite%
{nielsen} non-Markovian channel arises by introducing two bipartite Lindblad
terms $[\alpha =x,y$ in Eq. (\ref{LindbladInteraction})] defined by the
operators $V_{x}=\sqrt{p}\sigma _{x}\otimes \sigma ,$ and $V_{y}=\sqrt{1-p}%
\sigma _{y}\otimes \sigma ,$ where the parameter $p$ satisfies$\ 0<p<1.$
With the same collision statistics [Eq. (\ref{WaiterFluor})], in this case
the stationary system state becomes $\rho _{\infty }^{s}=(1/2)\mathrm{I}_{s}.
$ A thermal stationary state can be obtained by considering a generalized
amplitude damping superoperator \cite{nielsen}.

%Taking into account the
%results of Ref. \cite{buzeMarkov} one may also consider interactions that
%can be read as two-qubit logical operations. A partial controlled-not
%operation lead to a decoherence channel, while a partial swap operations may
%induce a dynamics equivalent to a zero-temperature reservoir.

\section{Generalized collisional models}

In the previous sections we associated the basic master equation of the
collision model [Eq. (\ref{MasterColision})] with an underlying Markovian
microscopic dynamics, Eq. (\ref{LindbladBipartito}). Furthermore, the
realizations of the model, given that the ancilla system is continuously
monitored in time, were established on the basis of the quantum jump
approach. In this section, we show that these results also apply in
different possible generalizations of the basic approach.

\subsection{Non-stationary renewal collision dynamics}

The basic ingredients of the present approach remain valid when the
evolution of the ancilla system, in the bipartite Lindblad dynamics (\ref%
{LindbladBipartito}), depends explicitly on time, $\mathcal{L}%
_{a}\rightarrow \mathcal{L}_{a}(t).$ Under this situation, the main change
is the measurement statistics. While it remains a renewal process, the
waiting time distribution explicitly depends on the observation time. This
case can be worked out with the elements introduced in the previous sections.

\subsection{Non-renewal collision statistics}

With the same system realizations, the formalism may becomes non-renewal
when the measurement process is non-renewal. Basically this situation occurs
whenever the ancilla resetting state is not always the same. This case
arises, for example, when the operators (\ref{Operators}) are generalized as%
\begin{equation}
V_{\alpha lk}=V_{\alpha }\otimes \left\vert a_{l}\right\rangle \left\langle
a_{\mathrm{0}}^{k}\right\vert .
\end{equation}%
Hence, instead of a unique state $\left\vert a_{\mathrm{0}}\right\rangle ,$
here many of them play the same role. Assuming that the measurement
apparatus is sensitive to \textquotedblleft all
transitions\textquotedblright\ $\left\vert a_{\mathrm{0}}^{k}\right\rangle
\rightsquigarrow \left\vert a_{l}\right\rangle ,$ the stochastic ancilla
becomes non-renewal. This case may corresponds, for example, to optical
cascade systems \cite{carmichael}.

While the structure of the systems realizations remains the same, the
statistics of the inter-event time intervals can only be determinate by
knowing the ancilla state at all times. Therefore, for generating the system
realizations unavoidably one also must to generates the ancilla realizations.

\subsection{Non-Markovian inter-collision dynamics}

Maintaining the renewal property, in Ref. \cite{VacCol} Vacchini introduced
an interesting generalization that consists in assuming that the inter-event
dynamics is non-Markovian. This situation naturally arises when considering
a system interacting successively with a string of qubits systems \cite%
{tanos,buzeta,Giova}.

Instead of the Markovian evolution defined by Eq. (\ref%
{IntereventSystemDynamics}), it is taken as%
\begin{equation}
\rho _{s}^{\mathrm{st}}(t)=\mathcal{G}(t-\tau )[\rho _{s}^{\mathrm{st}}(\tau
)],  \label{GDefinition}
\end{equation}%
where $\mathcal{G}(t)$ is an arbitrary (trace preserving) completely
positive propagator that cannot be written as a semigroup, $\mathcal{G}%
(t)\neq \exp [t\mathcal{L}_{s}]$ \cite{VacCol}. Here, we demonstrate that
this case can also be covered with the present formalism.

The generalized QCM can be embedded in a tripartite underlying Lindblad
equation. Hence, besides the system of interest $S,$ the ancilla system $A,$
we consider an extra auxiliary system $B.$ The evolution of their joint
density matrix $\rho _{t}^{sab}$ is written as%
\begin{equation}
\frac{d}{dt}\rho _{t}^{sab}=\mathcal{L}\rho _{t}^{sab}=(\mathcal{L}_{sb}+%
\mathcal{L}_{a}+\mathcal{C}_{sab})\rho _{t}^{sab}.
\label{LindbladTripartito}
\end{equation}%
The first superoperator reads%
\begin{equation}
\mathcal{L}_{sb}=\mathcal{L}_{s}+\mathcal{L}_{b}+\mathcal{C}_{sb}.
\end{equation}%
Here, $\mathcal{L}_{s}$ and $\mathcal{L}_{b}$\ are arbitrary Lindblad
equations for the systems $S$ and $B$ respectively. $\mathcal{C}_{sb}$ is an
extra Lindblad contribution that introduce an arbitrary interaction (unitary
and dissipative) between them. As before, $\mathcal{L}_{a}$ defines the
dynamics of the ancilla system $A.$ The contribution $\mathcal{C}_{sab}$
introduces a dissipative interaction between the three systems,%
\begin{equation}
\mathcal{C}_{sab}[\rho ]=\sum_{\alpha ,l,m}\gamma _{l}([V_{\alpha lm},\rho
V_{\alpha lm}^{\dag }]+[V_{\alpha lm}\rho ,V_{\alpha lm}^{\dag }]),
\label{tripartitaLindblad}
\end{equation}%
where $\gamma _{l}$ are the dissipative rates and the operators\ are%
\begin{equation}
V_{\alpha lm}=V_{\alpha }\otimes \left\vert a_{l}\right\rangle \left\langle
a_{\mathrm{0}}\right\vert \otimes \left\vert b_{\mathrm{0}}\right\rangle
\left\langle b_{m}\right\vert .  \label{TripartiteOperator}
\end{equation}%
The system operators $V_{\alpha }$ and the states $\left\vert
a_{l}\right\rangle $ are the same than in Eq. (\ref{Operators}), where the
index $l=1,2,\cdots ,\dim \{\mathcal{H}_{a}\}-1$ does not include the single
state $\left\vert a_{\mathrm{0}}\right\rangle .$ On the other hand, the
states $\left\vert b_{m}\right\rangle ,$ $m=0,1,\cdots \dim \mathcal{H}%
_{b}-1 $ form a complete orthonormal basis in the Hilbert space $\mathcal{H}%
_{b}$\ of $B.$ Notice that here the state $\left\vert b_{\mathrm{0}%
}\right\rangle $ must be included in the summation index $m.$ For
simplicity, the tripartite initial state is chosen separable%
\begin{equation}
\rho _{0}^{sab}=\rho _{0}^{s}\otimes \bar{\rho}_{a}\otimes \left\vert b_{%
\mathrm{0}}\right\rangle \left\langle b_{\mathrm{0}}\right\vert ,
\label{CITripartita}
\end{equation}%
where $\rho _{0}^{s}$ is an arbitrary system state and $\bar{\rho}_{a}$
follow from Eq. (\ref{AncillaReset}).

We determine the system realizations over the basis of a standard quantum
jump approach formulated on the basis of Eq. (\ref{LindbladTripartito}). As
before, the measurement apparatus is only sensitive to transitions of the
auxiliary system $A.$ Therefore, the transformation associated to each
detection event, instead of Eq. (\ref{MBpartito}), here reads%
\begin{equation}
\mathcal{M}\rho =\frac{\mathcal{E}_{s}[\sum_{m}\left\langle a_{\mathrm{0}%
}b_{m}\right\vert \rho \left\vert a_{\mathrm{0}}b_{m}\right\rangle ]}{%
\mathrm{Tr}_{s}[\sum_{m}\left\langle a_{\mathrm{0}}b_{m}\right\vert \rho
\left\vert a_{\mathrm{0}}b_{m}\right\rangle ]}\otimes \bar{\rho}_{a}\otimes
\left\vert b_{\mathrm{0}}\right\rangle \left\langle b_{\mathrm{0}%
}\right\vert .  \label{MTripartito}
\end{equation}%
The collisional superoperator $\mathcal{E}_{s}$ is given by Eq. (\ref%
{Esuperoperator}). On the other hand, the (tripartite) conditional dynamics
can be written as in Eqs. (\ref{Conditional}) and (\ref{UnormalizadaT}).
Nevertheless, here the superoperator $\mathcal{D}$ reads%
\begin{equation}
\mathcal{D}\rho =(\mathcal{L}_{sb}+\mathcal{L}_{a})\rho -\frac{\gamma }{2}%
\{\left\vert a_{\mathrm{0}}\right\rangle \left\langle a_{\mathrm{0}%
}\right\vert ,\rho \}_{+}.  \label{DTripartito}
\end{equation}%
In deriving this result we used that $\sum_{\alpha }V_{\alpha }^{\dag
}V_{\alpha }=\mathrm{I}_{s},$ and $\sum_{b=0}^{\dim \mathcal{H}%
_{b}-1}\left\vert b_{m}\right\rangle \left\langle b_{m}\right\vert =\mathrm{I%
}_{b}.$ With the previous definition of $\mathcal{D},$ the expression for
the survival probability, Eq. (\ref{SurvivalBipartita}), remains almost the
same, $P_{0}(t-\tau |\rho )=\mathrm{Tr}_{sab}[e^{t\mathcal{D}}\rho ].$

Over the basis of the previous two equations and the initial condition (\ref%
{CITripartita}), it is simple to conclude that the tripartite stochastic
state $\rho _{sab}^{\mathrm{st}}(t)$ $[\overline{\rho _{sab}^{\mathrm{st}}(t)%
}=\rho _{t}^{sab}]$ can be written at all times as%
\begin{equation}
\rho _{sab}^{\mathrm{st}}(t)=\rho _{sb}^{\mathrm{st}}(t)\otimes \rho _{a}^{%
\mathrm{st}}(t).  \label{SeparadorRhoEstocastica}
\end{equation}%
The dynamics for the ancilla state $\rho _{a}^{\mathrm{st}}(t)$ remains the
same as before, that is, Eqs. (\ref{AncillaTransformation}) to (\ref%
{DAncilla}) are not modified by the introduction of system $B.$ In
consequence, the measurement statistics, defined by the survival probability
(\ref{SurvivalCYQRW}), or equivalently the waiting time distribution (\ref%
{waiting}), is also the same.

The induced stochastic system dynamics follows from $\rho _{s}^{\mathrm{st}%
}(t)=\mathrm{Tr}_{ab}[\rho _{sab}^{\mathrm{st}}(t)].$ Hence, in each
recording event the state suffer the disruptive transformation%
\begin{equation}
\rho _{s}^{\mathrm{st}}(t)\rightarrow \mathrm{Tr}_{ab}[\mathcal{M}\rho
_{sab}^{\mathrm{st}}(t)]=\mathcal{E}_{s}[\rho _{s}^{\mathrm{st}}(t)].
\label{TripartoMeasurement}
\end{equation}%
This expression follows straightforwardly from Eq. (\ref{MTripartito}),
after using Eq. (\ref{SeparadorRhoEstocastica}) and noting that $%
\sum_{m}\left\langle b_{m}\right\vert \bullet \left\vert b_{m}\right\rangle =%
\mathrm{Tr}_{b}[\bullet ].$ On the other hand, the inter-collision dynamic
[Eq. (\ref{IntereventSystemDynamics})], here is $\rho _{s}^{\mathrm{st}}(t)=%
\mathrm{Tr}_{ab}[\mathcal{T}_{c}(t-\tau )\rho _{sab}^{\mathrm{st}}(\tau )].$
Given that $\mathcal{T}_{c}(t)$ follows from Eqs. (\ref{Conditional}) and (%
\ref{UnormalizadaT}), the operator $\mathcal{D}$ [Eq. (\ref{DTripartito})]
and the separability property defined by Eqs. (\ref{MTripartito}) and (\ref%
{SeparadorRhoEstocastica}) lead to%
\begin{equation}
\rho _{s}^{\mathrm{st}}(t)=\mathrm{Tr}_{b}\{\exp [(t-\tau )\mathcal{L}%
_{sb}]\left\vert b_{\mathrm{0}}\right\rangle \left\langle b_{\mathrm{0}%
}\right\vert \}\rho _{s}^{\mathrm{st}}(\tau ).
\end{equation}%
This conditional dynamics recovers the phenomenological proposal Eq. (\ref%
{GDefinition}). Hence, the non-Markovian propagator $\mathcal{G}(t)$ reads%
\begin{equation}
\mathcal{G}(t)=\mathrm{Tr}_{b}[\exp (t\mathcal{L}_{sb})\left\vert b_{\mathrm{%
0}}\right\rangle \left\langle b_{\mathrm{0}}\right\vert ].
\label{GMicrosocopico}
\end{equation}%
This is the main result of this section. It implies that the generalized
phenomenological approach of Ref. \cite{VacCol}\ can be described over the
basis of a tripartite Markovian evolution. If $\mathcal{L}_{sb}=\mathcal{L}%
_{s}+\mathcal{L}_{b},$ that is, when the system $S$ and the ancilla $B$ do
not interact, the formalism of the previous section, $\mathcal{G}(t)=\exp (t%
\mathcal{L}_{s}),$ is recovered. Hence, given the structure of the operators
(\ref{TripartiteOperator}), it becomes clear that the main role of system $B$
is to modify the inter-collision system dynamics.

The realizations defined by the measurement transformation (\ref{MTripartito}%
) and the inter-event dynamics (\ref{GMicrosocopico}) are similar to that
found in Ref. \cite{OneChannel}, where a non-Markovian generalization of the
quantum jump approach was defined over a similar basis by assuming that the
system of interest is submitted to a measurement process. Nevertheless, the
present treatment explicitly demonstrate that collisional dynamics can only
be linked with a quantum measurement theory if the monitoring action is
performed over the auxiliary ancilla system.

The non-local character of the propagator $\mathcal{G}(t)$ can be showed by
writing Eq. (\ref{GMicrosocopico}) in the Laplace domain as $\mathcal{G}(u)=%
\mathrm{Tr}_{b}[(u-\mathcal{L}_{sb})^{-1}\left\vert b_{\mathrm{0}%
}\right\rangle \left\langle b_{\mathrm{0}}\right\vert ].$ This expression
can be rewritten as $\mathcal{G}(u)=\{\mathrm{Tr}_{a}[(u-\mathcal{L}%
_{sb})^{-1}(u-\mathcal{L}_{sb})\left\vert b_{\mathrm{0}}\right\rangle
\left\langle b_{\mathrm{0}}\right\vert ]\}^{-1}\times \{[\mathcal{G}%
(u)]^{-1}\}^{-1}.$ Using in the curly brackets that $X^{-1}\times
Y^{-1}=(Y\times X)^{-1},$ where $X$ and $Y$ are arbitrary matrices, it
follows $\mathcal{G}(u)=\{[\mathcal{G}(u)]^{-1}(u\mathrm{Tr}_{a}[(u-\mathcal{%
L}_{sb})^{-1}\left\vert b_{\mathrm{0}}\right\rangle \left\langle b_{\mathrm{0%
}}\right\vert ]-\mathrm{Tr}_{a}[(u-\mathcal{L}_{sb})^{-1}\mathcal{L}%
_{sb}\left\vert b_{\mathrm{0}}\right\rangle \left\langle b_{\mathrm{0}%
}\right\vert ])\}^{-1},$ which in turn leads to%
\begin{equation}
\mathcal{G}(u)=\frac{1}{u+\mathcal{K}(u)},
\end{equation}%
where the\ system superoperator $\mathcal{K}(u)$ is%
\begin{equation}
\mathcal{K}(u)\!=\!\Big{\{}\!\mathrm{Tr}_{b}\!\Big{[}\!\frac{1}{u-\mathcal{L}%
_{sb}}\!\left\vert b_{\mathrm{0}}\right\rangle \left\langle b_{\mathrm{0}%
}\right\vert \!\Big{]}\!\Big{\}}^{-1}\!\mathrm{Tr}_{b}\!\Big{[}\!\frac{1}{u-%
\mathcal{L}_{sb}}\!\mathcal{L}_{sb}\!\left\vert b_{\mathrm{0}}\right\rangle
\left\langle b_{\mathrm{0}}\right\vert \!\Big{]}.  \notag
\end{equation}%
Hence, in the time domain we get%
\begin{equation}
\frac{d}{dt}\mathcal{G}(t)=\int_{0}^{t}dt^{\prime }\mathcal{K}(t-t^{\prime })%
\mathcal{G}(t^{\prime }),  \label{d/dtT(t)}
\end{equation}%
where $\mathcal{K}(t-t^{\prime })$ is defined by its Laplace transform $%
\mathcal{K}(u).$

The evolution of $\rho _{t}^{s}$ can be obtained from Eq. (\ref%
{LindbladTripartito}) by using projector techniques. A simpler way is to
calculate the average behavior of the ensemble of stochastic realizations
(see Ref. \cite{VacCol}). On the other hand, the QCM introduced by
Ciccarello, Palma, and Giovannetti in Ref. \cite{tanos}, which relies on
interaction with a qubits-string, can also be recovered from the present
approach. In fact, as demonstrated in Ref. \cite{VacCol} it arises by taking 
$\mathcal{E}_{s}\rightarrow \mathrm{I}_{s}.$ Hence, each collision only
resets the evolution induced by $\mathcal{G}(t).$ The results presented by
Rybar \textit{et. al. }in Ref. \cite{buzeta} rely on a similar approach. All
non-Markovian effects arise because the ancilla string begin in a correlated
state \cite{Giova}. Nevertheless, in our approach that formalism seems to be
equivalent to a system-ancilla dynamics coupled via a unitary evolution,
which in turn leads to a random-like superposition of Hamiltonian system
propagators. Therefore, extra analysis are necessary for establishing a full
mapping between both approaches.

In what follows we analyze how different underlying dynamics lead to
dephasing and dissipative inter-collision dynamics \cite{VacCol,tanos}.

\subsubsection{Dephasing inter-collision dynamics}

In this example, both the system an the ancillas are two-level systems.
Their tripartite Markovian evolution is given by Eq. (\ref%
{LindbladTripartito}). In an interaction representation with respect to the
system Hamiltonian, we write%
\begin{equation}
\mathcal{L}_{sb}[\rho ]=\frac{-i}{\hbar }[H_{sb},\rho ]=\frac{-i\lambda }{2}%
[\sigma _{z}\otimes \mathrm{I}_{a}\otimes \sigma _{x},\rho ],
\label{LsbExample}
\end{equation}%
where $\sigma _{j},$ $j=x,y,z,$ are the Pauli matrices defined in each
Hilbert space. Hence, the system of interest $S$ and the auxiliary system $B$
are coupled via a Hamiltonian interaction. The isolated dynamics of ancilla $%
A$ is unitary%
\begin{equation}
\mathcal{L}_{a}[\rho ]=\frac{-i}{\hbar }[H_{a},\rho ]=\frac{-i\Delta }{2}[%
\mathrm{I}_{s}\otimes \sigma _{x}\otimes \mathrm{I}_{b},\rho ].
\label{UnitaryAncilla}
\end{equation}%
The dissipative tripartite interaction [Eq. (\ref{tripartitaLindblad})] reads%
\begin{equation}
\mathcal{C}_{sab}[\rho ]=\gamma \sum_{m=0,1}([V_{m},\rho V_{m}^{\dag
}]+[V_{m}\rho ,V_{m}^{\dag }]).
\end{equation}%
The index $m=0,1,$ runs over the basis $\{\left\vert b_{\mathrm{0}%
}\right\rangle ,\left\vert b_{\mathrm{1}}\right\rangle \}$\ of system $B.$
The two operators $V_{m}$ are%
\begin{equation}
V_{m}=\sigma _{x}\otimes \sigma \otimes \left\vert b_{\mathrm{0}%
}\right\rangle \left\langle b_{m}\right\vert ,  \label{VmTripartito}
\end{equation}%
where as before $\sigma $ is the lowering operator, here defined in the
Hilbert space of system $A.$ Consistently with Eq. (\ref{CITripartita}), the
initial tripartite state is%
\begin{equation}
\rho _{0}^{sab}=\rho _{0}^{s}\otimes \left\vert -\right\rangle \left\langle
-\right\vert \otimes \left\vert b_{\mathrm{0}}\right\rangle \left\langle b_{%
\mathrm{0}}\right\vert .
\end{equation}

From the previous definitions, Eqs. (\ref{MTripartito}) and (\ref%
{TripartoMeasurement}) lead to the collision system superoperator%
\begin{equation}
\mathcal{E}_{s}[\rho ]=\sigma _{x}\rho \sigma _{x}.  \label{SuperX}
\end{equation}%
Notice that $\sigma _{x}$ arise from the first (system) operator
contribution in Eq. (\ref{VmTripartito}). On the other hand, the dynamics of
system $A$ again is defined by Eq. (\ref{AncillaFluor}). Consequently, the
waiting time distribution is given by Eq. (\ref{WaiterFluor}). The
inter-collision dynamic follows from Eq. (\ref{GMicrosocopico}) and (\ref%
{LsbExample}). By an explicit calculation, we get the completely positive
(non-Markovian) dephasing superoperator%
\begin{equation}
\mathcal{G}(t)\rho =\frac{1}{2}[1+d(t)]\rho +\frac{1}{2}[1-d(t)]\sigma
_{z}\rho \sigma _{z},  \label{CosPropa}
\end{equation}%
which in turn can be rewritten as%
\begin{equation}
\mathcal{G}(t)\rho =\left( 
\begin{array}{cc}
\left\langle +\right\vert \rho \left\vert +\right\rangle & d(t)\left\langle
+\right\vert \rho \left\vert -\right\rangle \\ 
d(t)\left\langle -\right\vert \rho \left\vert +\right\rangle & \left\langle
-\right\vert \rho \left\vert -\right\rangle%
\end{array}%
\right) .
\end{equation}%
The function $d(t)$ defines the system coherences behavior. It reads $%
d(t)=\cos (\lambda t).$

In the first example worked out in Ref. \cite{VacCol}, the superoperator is
given by Eq. (\ref{SuperX}), while the propagator $\mathcal{G}(t)$ is given
by Eq. (\ref{CosPropa}) (see supplemental material of \cite{VacCol}). Hence,
our results provides a clear microscopic description for that
phenomenological model. The waiting time distribution, instead of Eq. (\ref%
{WaiterFluor}), is a classcial one like Eq. (\ref{WaiterClasica}). That case
can be recovered replacing the ancilla dynamics (\ref{UnitaryAncilla}) by%
\begin{equation}
\mathcal{L}_{a}[\rho ]=\beta ([A,\rho _{t}^{sa}A^{\dagger }]+[A\rho
_{t}^{sa},A^{\dagger }]),
\end{equation}%
with the operator%
\begin{equation}
A=\mathrm{I}_{s}\otimes \sigma ^{\dagger }\otimes \mathrm{I}_{b}.
\end{equation}%
As explained previously, diverse \textquotedblleft underlying
classical\textquotedblright\ waiting time distributions can be obtained by
adding extra ancilla states, all of then coupled by incoherent transitions.

\subsubsection{Dissipative inter-collision dynamics}

Instead of the dephasing evolution (\ref{CosPropa}), the inter-collision
dynamics may also lead to dissipative effects. This property is defined by
the superoperator $\mathcal{L}_{sb}$ [Eq. (\ref{LsbExample}) in the previous
example]. For example $\mathcal{L}_{sb}$ may correspond to a Jaynes-Cumming
interaction, which couples the system to a set of Bosonic field modes
initially in the vacuum state \cite{breuerbook,Entropy}. This case, which
has been studied in Refs. \cite{VacCol,tanos} can be analyzed over the basis
developed previously.

%Finally, let remark that from our approach some extra generalization of the
%collision model of Ref. \cite{VacCol} case may be formulated. For example,
%if system $B$ also affects the measurement transformation, a correlation
%between the events statistics and the inter-collision evolution may be
%introduced.

\section{Summary and conclusions}

%The main result of this paper is the finding of a clear and well defined
%physics basis for describing phenomenological collisional models.

Phenomenological QCMs provided an important theoretical tool for
establishing and describing non-Markovian completely positive dynamics. In
this paper we have developed a solid physics basis for understanding this
approach. It relies on a Markovian embedding of the non-Markovian system
density matrix evolution, which in turn allows to derive the
phenomenological trajectories from a quantum measurement theory.

%In fact, it was derived over the basis of Markovian Lindblad equations and the standard quantum jump approach.

First, we focused our analysis on the leading case in which the collision
statistics is defined by a renewal process, while the inter-event dynamics
is defined by a Markovian quantum semigroup. By using projector techniques
we demonstrated that the non-Markovian density matrix evolution [Eq. (\ref%
{MasterColision})] can be obtained, without involving any approximation,
from a bipartite Markovian dynamics where the system of interest interact
with an auxiliary ancilla system, Eq. (\ref{LindbladBipartito}). The memory
kernel that determines the system evolution becomes defined by the ancilla
dynamics, Eq. (\ref{LindbladKernel}). The proposed Markovian embedding
allows to associate a clear microscopic dynamics to the QCM. In fact,
Lindblad equations are linked with well defined microscopic dynamics.

In a second step, we assumed that the ancilla system is continuously
monitored in time. Hence, over the basis of the quantum jump approach
formulated for the bipartite dynamics, we find that the realizations of the
QCM are recovered from the marginal conditional stochastic system dynamics,
Eq. (\ref{RhoAverage}). In fact, each recording event of the ancilla
measurement apparatus lead to the collisional transformations of the
phenomenological approach, Eq. (\ref{SystemTransformation}). The
inter-collision system dynamics follows from the conditional bipartite
dynamics between detection events, Eq. (\ref{IntereventSystemDynamics}). The
waiting time distribution of the inter-event time interval also becomes
defined by the ancilla dynamics, Eq. (\ref{waiting}). In this way, the
phenomenological realizations of the collisional approach were derived from
a quantum measurement theory.

The Markovian embedding and the link with the quantum jump approach were
explicitly shown through an example where the dynamics of both the system of
interest and the auxiliary one develop in two dimensional Hilbert spaces
(Figs. 1 and 2). In contrast to phenomenological formulations, here the
collision statistics arises from quantum coherent effects developing in the
ancilla Hilbert space. A system-to-environment back flow of information
characterize the dynamics. In contrast, when the ancilla dynamic is
completely incoherent, this feature is absent.

%In spite of the simplicity of the model, the present approach allowed us to
%demonstrate that even the simple evolutions, where the system dynamics is
%frozen between collisional events, can lead to non-Markovian effects such as
%an environment-to-system back flow of information (Fig. 2).

The previous finding provide a solid basis for proposing different
generalizations of the QCM. For example, non-stationary renewal collision
dynamics can be obtained by introducing an explicit time dependence in the
ancilla dynamics. Non-renewal collision statistics can be related to a
non-renewal ancilla measurement process. On the other hand, we showed that
by introducing a second auxiliary system the inter-collision dynamics
becomes defined by a non-Markovian propagator, Eq. (\ref{GMicrosocopico}).
This finding allowed us to recover a recent proposed generalization of the
QCM \cite{VacCol}, which in fact can also be embedded in a Markovian
evolution and their realizations derived from a quantum measurement theory.
From this result, we also concluded that some non-Markovian collisional
models formulated in terms of qubits logical operations \cite{tanos} can
also be recovered from our formalism.

The present analysis allow us to read the phenomenological QCMs from a novel
perspective. Besides a solid physical basis of the corresponding
non-Markovian dynamics, the developed approach provides an alternative and
power tool for describing non-Markovian memory effects in open quantum
systems.

% On the other hand, based on Ref. \cite{buzeta}, an extended
%analysis of the developed approach in presence of initial-system-ancilla
%correlations may be of interest when characterizing non-Markovian qubits
%dynamics.

\section*{Acknowledgments}

This work was supported by CONICET, Argentina, under Grant No. PIP
11420090100211.

\end{document}